\documentclass[12pt]{article}
\usepackage{a4}
\usepackage{epsf}
\usepackage{rotating}
\usepackage{epsfig}
\usepackage{amssymb}
\usepackage{colortbl}
\usepackage{colordvi}

\newcommand{\beq}{\begin{eqnarray}}
\newcommand{\eeq}{\end{eqnarray}}

\newcommand{\Gev}{{\rm GeV}}
\newcommand{\Mev}{{\rm MeV}}
\newcommand{\be}{\begin{equation}}
\newcommand{\ee}{\end{equation}}
\newcommand{\lwrsim}{\raise0.3ex\hbox{$<$\kern-0.75em\raise-1.1ex\hbox{$\sim$}}}

\newcommand \VEV [1] {\left\langle{#1}\right\rangle}
\newcommand{\rhob}{\overline{\rho}}
\newcommand{\rhodb}{\overline{\rho^2}}

\def\C2#1#2{({\cal C}_2)_{#1}^{#2}}

\def\eq#1{eq. (\ref{#1})}

\begin{document}
\setcounter{page}{1}
\begin{flushright}
\begin{tabular}{l}
{\tt LPT Orsay 03-104}\\
{\tt UHU-FT/03-23 }\\
{\tt FAMN SE-02/04}\\
\end{tabular} 
\end{flushright}
\begin{center}
\bf{\huge Modified instanton profile effects from lattice Green functions.}\\
\end{center}  
\vskip 0.8cm

\begin{center}{\bf  Ph. Boucaud$^a$, F. De Soto$^{b}$, A. Le Yaouanc$^a$, J.P. Leroy$^a$, 
J. Micheli$^a$,
O. P\`ene$^a$, J. Rodr\'{\i}guez--Quintero$^c$   }\\
\vskip 0.5cm 
$^{a}$ {\sl Laboratoire de Physique Th\'eorique~\footnote{Unit\'e Mixte 
de Recherche du CNRS - UMR 8627}\\
Universit\'e de Paris XI, B\^atiment 210, 91405 Orsay Cedex,
France}\\
$^b${\sl Dpto. de F\'{\i}sica At\'omica, Molecular y Nuclear \\
Universidad de Sevilla, Apdo. 1065, 41080 Sevilla, Spain} \\
$^c${\sl Dpto. de F\'{\i}sica Aplicada \\
Fac. Ciencias Experimentales, Universidad de Huelva, 21071 Huelva, Spain} \\
\end{center}

\medskip
\begin{abstract}
We trace here instantons through the 
analysis of pure Yang-Mills gluon Green functions in the Landau gauge for a window of IR momenta 
($0.4$~GeV $< k < 0.9$~GeV). 
We present lattice results that can be fitted only after substituting the BPST profile in the 
Instanton liquid model (ILM) by one based on the Diakonov and Petrov variational methods. 
This also leads us to gain information on the parameters of ILM. 
\end{abstract}

\section{Introduction}

An appealing approach to analytically understand some of the
non-perturbative features of QCD is the evaluation of
quantum fluctuations around topologically non-trivial classical
solutions through the expansion of the path integral around these
solutions. In fact these considerations are often generalized 
to configurations which are not exact solutions of the field equations but 
close to them and which we will name quasi-classical field configurations. 

Famous examples of non-trivial solutions of classical equations of motion
are instantons~\cite{Belavin:fg,'tHooft:fv}. 
 Quasi-classical solutions considered in instanton 
 liquid models~\cite{ILM,Diakonov:1983hh}  
provide a successful connection between the instanton zero modes and the QCD 
chiral symmetry breaking (See ref. \cite{Schafer:1996wv} for a 
good review on the subject). 

More recently, it has been proven in ref.~\cite{Faccioli:2003qz} that 
instanton model predictions for quark-quark interaction agree with 
non-perturbative QCD lattice results better than those 
from Schwinger-Dyson (SD) models with a perturbative structure of the 
QCD interaction (vector quark-gluon coupling parametrization). 
In general, a dominance of instanton-induced effects on the dynamics of the  
QCD light-quark sector seems to emerge, although it is not excluded that 
SD models with pseudo-scalar and scalar quark-gluon couplings might capture 
this emerging instanton physics.
On the other hand, despite instantons have been first considered as a possible
explanation of confinement \cite{Polyakov:rs}, it is now generally
accepted that they do not generate the area law for Wilson
loops~\footnote{Another set of solutions of classical field equations,
merons~\cite{Callan:qs}, has been recently reexamined as candidates to
explain confinement~\cite{Lenz:2003jp}.}. 

We have  recently argued~\cite{Boucaud:2002fx} that instantons, or
instanton-like structures, have dramatic effects on the low momentum Green
function in Yang-Mills  theories and that they can explain the observed 
$\propto k^4$ behaviour of the non-perturbative MOM QCD coupling 
constant computed on the lattice.
Notice that this remark, not only advocates in favor of the presence
of these quasi-classical structures in the lattice gauge configurations, 
but also indicates that the quantum fluctuations do not contribute 
significantly to the Green functions in this momentum regime. In the present paper 
we try a low momentum description of two- and three-gluon Green functions through Instanton 
liquid model (ILM).

The succesful description of the MOM QCD coupling constant in the 
low-momentum regime is based on the sum-ansatz approach, that builds the 
classical solution as a linear combination of modified instantons. 
Although instanton profile modifications play no role 
in obtaining the coupling constant, at least in the first 
approximation, any description of 
two- and three-gluon Green functions makes mandatory to further elaborate on 
the nature of this modifications. To this goal we will follow the Diakonov \& 
Petrov (DP) sum-ansatz approach~\cite{Diakonov:1983hh}. As will be discussed later,
 different aspects of 
this approach have been criticised by Shuryak and Verbaarshot, but it provides us 
with a framework able to estimate instanton effects through 
analytical or semi-analytical computations which seem to work reasonably well. This is 
the ``phenomenological'' point of view which we adopt in this paper to extract some
understanding of the low-momentum behaviour of lattice Green functions.

The paper is organized in six sections. In section 2 we discuss the pattern of the running 
with momenta of the QCD coupling constant. Section 3 is devoted to study the instanton 
profile modification within the DP approach. In section 4, we show that the low momentum 
behaviour of lattice gluon Green functions is rather well described by ILM only after including 
instanton profile modifications and discuss how the large instanton density obtained from the 
fits is expected to be reduced by light dynamical quarks in full QCD. 
In section 5, the effect of instanton radius distribution is discussed. We finally 
conclude in section 6. 

At the end of the day, of course, rather important questions still remain open: 
why do quantum effects appear to be
suppressed in this low momentum range ? Has this some connection
with confinement ?

\section{The QCD coupling constant: four regimes}
\label{4-regimes}

We have presented  in Ref. \cite{Boucaud:2002fx} a preliminary 
claim of instanton dominance at low energy by analyzing 
in Landau gauge the following ratio of pure Yang-Mills Green functions:
\beq
\alpha_s(k^2) \ = \ \frac 1 {4 \pi} \left( \frac{G^{(3)}(k^2,k^2,k^2)}
{(G^{(2)}(k^2))^3} Z_{\rm MOM}^{3/2}(k^2) \right)^2 \ = \frac{k^6}{4 \pi} \ 
\frac{\left( G^{(3)}(k^2,k^2,k^2) \right)^2}{\left(G^{(2)}(k^2)
\right)^3} ,
\label{mom}
\eeq
which is a non-perturbative MOM definition of the coupling constant, 
where $G^{(n)}$ is the gluon n-point correlation function and 
$Z_{\rm MOM}=k^2 G^{(2)}$ is the gluon propagator renormalization
constant in  MOM scheme~\footnote{We keep the name ``coupling
constant'' for this well defined quantity although it could be argued
that this name is not really appropriate in the low momentum
regime.}.

\begin{figure}
\begin{center}
\begin{tabular}{c}
\includegraphics[width=25pc]{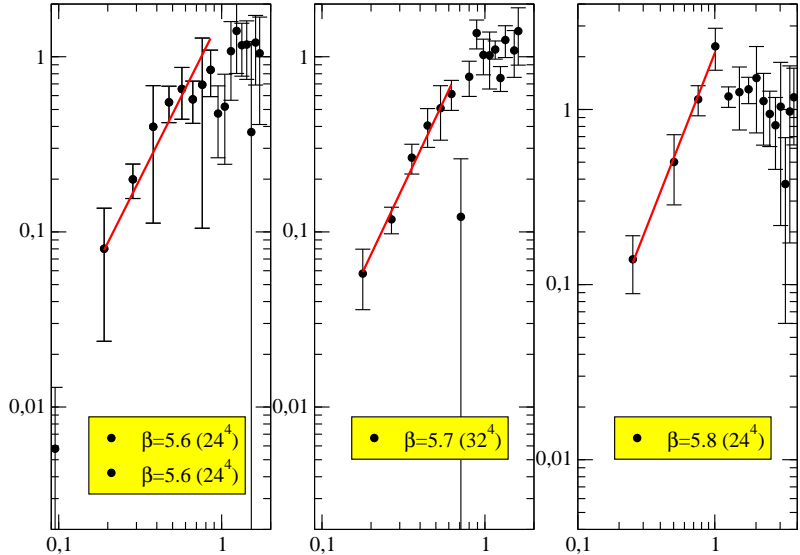} 
\end{tabular}
\caption{\small{\it The low momentum $k^4$ behavior of
$\alpha_s(k^2)$ on log-log plots for three  different lattices
((5.6,24), (5.7,32), (5.8,24)) using 20 configurations.  
Notice that  above $\sim 1$ GeV, no smooth line can be easily
drawn  joining the lattice data.}} 
\label{Fig1}
\end{center}
\end{figure}

In Fig. \ref{Fig1} we show on a log-log plot, that a roughly-$k^4$
power  law is satisfied by the  lattice evaluations of
$\alpha_s(k^2)$, eq. (\ref{mom}), up to 0.8-0.9 GeV, for three
different lattices, strongly supporting a  quasi-classical
description \cite{Boucaud:2002fx}.

A very striking feature of the results  shown in Fig. \ref{Fig1},
obtained with a low statistics of 20 configurations, is that there is
a rather sharp transition at $\sim 1 \Gev$ between two regimes: below this
scale $\alpha_s(k^2)$ does not seem to fluctuate much, it agrees well,
with a small $\chi^2$, with the expected $k^4$
linear behaviour in spite of the small statistics and it complies with an
instanton-like picture, while above
that scale the data suddenly  deviate from the $k^4$ law and become
apparently fuzzier~\footnote{  The smooth curve shown in  fig 1(b)
in~\cite{Boucaud:2002fx} has been reached with  1000
configurations.}, a fuzziness which is confirmed by the larger $\chi^2$
which affects any smooth fit.  A tempting interpretation is that this 
fuzziness has to do with a strong influence of quantum fluctuations which
increase suddenly above 1 GeV. We may understand this as follows: for
a given profile and a given  instanton radius, the ILM predicts  no
statistical fluctuation of the Green functions. The average over the 
instantons locations and their color orientation does not create any
noise on Green functions which are translational and color rotation
invariant. Only the dispersion (to be studied at length in this
paper) of the instanton radius as well as varying effects of
the neighbouring instantons produces some statistical noise.  On the
contrary the quantum contributions to Green function are generated by 
statistical fluctuations of the gauge fields around zero and Green
functions appear as correlations in this statistical system. 
Some confirmation of this interpretation results of the analysis of 
the same gluon Green functions after applying a {\it cooling} procedure 
that kills short-distance (quantum) correlations~\cite{cooling}.

Altogether we may distinguish four regimes:
\begin{itemize}
\item Above 2.6 GeV we have shown~\cite{Boucaud:2000nd,Unquenched} that the
lattice data were dominated by perturbative QCD with a significant
non-perturbative  correction describable via OPE by the expectation
value of $A^2$. This means clearly a dominance of quantum
fluctuations with small corrections from the $A^2$ condensate which
may be generated by the quasi-classical solutions. Indeed the
quasi-classical solutions, being large structures,  are seen  by the
hard propagating gluons as  an  effectively translational invariant
background. It is then easy to show that their  dominant effect is 
amenable to an OPE treatment of the lowest dimension  operator: 
$A^2$~\cite{Boucaud:2002nc}.   
\item Between $\sim$ 0.4 and $\sim$ 0.9 GeV the quasi-classical contributions
dominate and the quantum effects are strongly depressed.
\item The 1.0 to 2.6 GeV region shows a strong quantum effect.  
However, it is not
at all describable in terms of perturbation theory. A description in
terms of quantum fluctuations in a quasi-classical background should
be tried. The latter background can no longer be treated as  simply as
in the large momentum regime.   Other non-perturbative effects may
also play a role, for example related to confinement.  
\item The very low momentum region below $\sim 0.4$ GeV is still
``terra incognita'' and, being of the order of $2\pi/L$, ($L$ being 
the lattice length) possibly strongly sensitive to finite volume artifacts.
\end{itemize} 

\section{A modified profile for the instanton-liquid model}

In the following, we will try to describe the Green functions
assuming that the relevant quasi-classical solutions are instanton
liquids.  Let us compute  the quantity in eq. (\ref{mom}). We
describe gluon fields in the low  energy regime as  a superposition of
modified instantons, a quasi-classical
background dominance being assumed. 
In this framework, the gauge field in Landau gauge will
be given by:

\beq g\, B_{\mu}^{a}({\bf x})= 2 \sum_i R^{ a \alpha}_{(i)}\,
\overline{\eta}^\alpha_{\mu \nu} \frac{(x_\nu-z_{\nu}^i)}{(x-z^i)^2}
\phi\left(\frac{|x-z^i|}{\rho_i}\right) \ , \label{amuins} \eeq 

\noindent where $z^i$ ($\rho_i$) are the center (radius) of the
instantons, $\overline{\eta}^\alpha_{\mu \nu}$ is known as 't Hooft
symbol, $R^{ a\, \alpha}_{(i)}$ are color rotations embedding the
canonical  $SU(2)$ instanton into the $SU(3)$ gauge group,
$\alpha=1,3$ ($ a=1,8$) is an $SU(2)$ ($SU(3)$)  color index, and 
the sum is extended over instantons and anti-instantons.

The classical solution for an isolated instanton is the standard BPST
one, $\phi(\xi)=1/(\xi^2+1)$~\cite{Belavin:fg}. Nevertheless, the
superposition (\ref{amuins}) with a BPST profile is not a solution of Yang-Mills
equations (since they are not linear). Hence, we will discuss below a
parameterization of the profile function inspired by an  approximated
minimization of the action for a finite density of 
instantons~\footnote{The function $\phi$ in eq. (\ref{amuins}) is related 
to the function P in eq. (6) of ref. \cite{Boucaud:2002fx} by the relation 
$P(\xi^2) = 2 \phi(\xi)/\xi^2$.}.

Then, just by assuming random color orientation and
instanton  position~\footnote{This means that we neglect the color
correlation which might exist, for example, between neighbouring
instantons, an assumption which is usually done and which amounts to
consider this instanton liquid as not being ordered.}, from \eq{amuins} 
we obtain (see \cite{Boucaud:2002fx}):

\beq\label{propins}
g^2 G^{(2)}_{(I)}(k^2)=  \frac n 8 <\rho^6  I(k\rho)^2> \nonumber \\
g^3 G^{(3)}_{(I)}(k^2,k^2,k^2)= \frac n {48\, k}<\rho^9  I(k\rho)^3>
\eeq
where $n$ stands for the instanton density, $<\cdots>$ denotes
average over the instanton radius for a given radius distribution
$\mu(\rho)$, normalized to 1, and where

\beq
I(s) \ = \frac{8 \pi^2} s\int_0^\infty\ z dz  J_2(sz)\,\phi(z) \ , \ \ \ \ s>0 \ ;
\label{Igen}
\eeq
$J_2$ being the  second order Bessel J~function. The factor $g^n$
 for n-points Green functions 
in l.h.s. of eq. (\ref{propins}) comes from the factor
 $g$ in the l.h.s. of eq.~(\ref{amuins}). 

Then, from Eq. (\ref{mom}) we get:
\beq\label{alpha}
\alpha_{s\,(I)}(k) \ = \ \frac{k^4}{18 \pi n} \ 
\frac{< \rho^9  I(k\rho)^3 >^2}
{< \rho^6  I(k\rho)^2 >^3}\ .
\eeq

As a first approximation, if we consider all instantons to have
 the same radius, we obviously obtain:

\beq\label{delta}
\frac{< \rho^9  I(k\rho)^3 >^2}
{< \rho^6  I(k\rho)^2 >^3} \ = \ 1 \ ,
\eeq
and recover an exact $k^4$-power law for any instanton profile.

In (\ref{delta}) the influence of the profile will only appear  as
a sub-leading contribution, that will also depend on the instanton
radius distribution (See section \ref{radii}.), while in (\ref{propins}) the leading 
contributions to the Green functions  depend on
the  profile and will therefore be used by us to gain some
understanding  about the instanton radial shape and radii. 
Indeed, it will be manifest (see fig. \ref{fits}) that the 
BPST profile cannot account 
for the low momentum behaviour of Green functions.

\paragraph{The variational Diakonov \& Petrov equation.\\ }

If the QCD vacuum can be understood as an instanton liquid of
finite density (See \cite{Schafer:1996wv}, for example), the BPST
profile  will no longer be valid (it is only a zero density
limit) especially at large distance from the instanton center 
where the overlap of neighboring instantons becomes important.
Being interested in the low momentum regime we cannot neglect 
these effects.

A possible method to include the effect of instanton interactions,
is to study the profile that minimizes the action of the instanton
ensemble. Such a procedure was used in \cite{Diakonov:1983hh}, where,
through the Feynman variational principle, the equation 
\beq\label{DPeq}
-\left( x^2 \frac{d}{dx^2} \right)^2 \phi +
\left( 1 +\frac{\alpha_{\rm DP}^2x^2}{4} \right)\phi - 3
\phi^2 + 2 \phi^3 + \frac {x^2} {6 \beta(\rho)} \frac 
{\delta C_{N_C}}{\delta \phi} = 0\ , 
\nonumber \\
\label{phi}
\eeq
was obtained for the best profile, $\phi$, eq. (6.7)
in~\cite{Diakonov:1983hh}~\footnote{We correct for a factor $x^2$
in the last term which misprinted in the quoted paper. The same for the square of 
the function $\phi$ in \eq{alDP}.}. 
In the last term $\beta(\rho)=8\pi^2/g^2(\rho)$ 
where $g(\rho)$ is the running coupling constant evaluated at the
instanton radius scale and $C_{N_C}$ is a factor containing the 
quantum corrections that multiply the
instanton liquid partition function and are basically defined by the
functional  determinants in eq. (2.19) of \cite{Diakonov:1983hh}.The
value of $\alpha_{\rm DP}$ which represents an average classical effect
of the other instantons on one of them is given by 
\beq\label{alDP}
\alpha_{\rm DP}^2= \frac 4 3 \gamma_0^2 n \frac{\beta(\overline \rho )}
{\beta(\rho)} < \int_0^\infty dx^2 \phi^2\left( \frac x \rho \right) > \ .
\eeq
where the parameter $\gamma_0$ is related to the instanton interaction strength,
\beq
\gamma_0^2 \ = \ \frac{27 N_c\pi^2}{4 (N_c^2-1)} \ .
\eeq
Thus, if we assume the same radius for all the instantons, this equation reduces to
\beq\label{baddens}
\alpha_{\rm DP}^2 \simeq \frac{9 N_c\pi^2}{N_c^2-1} n \int_0^\infty
dx^2 \phi^2\left(\frac x \rho \right) \ .
\eeq
The last term in eq. (\ref{phi}), involving a functional derivative of the
functional determinants, is hard to compute and explicitely violates scale invariance
(in the sense that $\phi(\frac{x}{a})$ is still a solution
if n is divided by $a^4$). Let us remark that this is true only if we do not take
into account the $\rho$ dependance of $\alpha_{\rm DP}$ due to
$\beta(\rho)$.

Now assuming that $\alpha_{\rm DP}$ is approximately constant, it is
argued in~\cite{Diakonov:1983hh}  (discussion before eq. (6.1)) that 
one can neglect this term to get a scale invariant equation governing the 
large distance shape of the profile function assumed to be dominated 
by classical interaction. Of course, we should borrow from the neglected term some scale 
invariance breaking to fix the instanton size and match to the BPST solution at
 small distances. This can be done 
{\it e.g.} by explicitely writing the profile as a function of $|x|/\rho$, 
$\rho$ defining the instanton size through the condition $\phi(1)=1/2$.

The term in $\alpha_{\rm DP}$ enforces a squeezing of the
instantons and therefore controls the large $x$ behaviour of
$\phi(x)$. For $|x| \gg \rho$ we can neglect in Eq. (\ref{phi})
the non linear terms obtaining the following Bessel equation:

\beq\label{Bessel}
\left\{ x^2 \frac {d^2} {d|x|^2} + |x| \frac d {d|x|} - 
\left( 4 + \alpha_{\rm DP}^2 x^2 \right) \ \right\} \ 
\phi\left(\frac{|x|}{\rho}\right) \ = \ 0 \ .
\eeq
Therefore we have:

\beq\label{K2}
\phi\left(\frac{|x|}{\rho}\right) 
\begin{array}{c} \rule[0.2cm]{0cm}{0.5cm} \sim \\ x \gg \rho \end{array}
c \ K_2(\alpha_{\rm DP} |x|) \  \ \sim \ \ c \ \frac{e^{-\alpha_{\rm DP} |x|}}{\sqrt{|x|}} \ ,
\eeq
where c is an unknown coefficient which could be determined
from the scale invariance breaking condition.
Let us make further remarks about Eq. (\ref{phi}) after neglecting the $C_{N_C}$ term:
the authors of ~\cite{Diakonov:1983hh} claim that at small x
the term proportional to $\alpha^2_{\rm DP}x^2$ becomes small and therefore
can be neglected, recovering the instanton equation. This term is nevertheless not
negligeable wih respect to $\phi -3 \phi^2 + 2 \phi^3$ which tends to 0
when $x\to 0$. Furthermore this equation has no solution going to 0 at
infinity and to 1 at $x = 0$ because a singularity emerges at some small finite
value of $x$.

\begin{figure}[h!]
\begin{center}
\includegraphics[width=20pc]{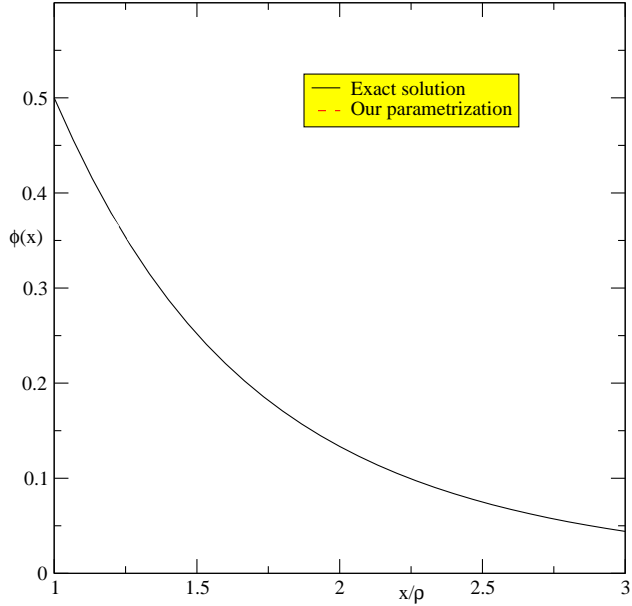}
\caption{\small \it Comparison between our parametrization of the profile
function wit $\alpha_{DP} \rho=0.68$ and the exact solution of the Diakonov and Petrov equation. }  
\label{Fig0}
\end{center}
\end{figure}

For the above reasons, we prefere to use in the next sections instead of direct solutions of \eq{Bessel} 
the following
parametrization:

\beq\label{para}
\phi\left(\frac{|x|}{\rho}\right) = 
\frac{(\alpha_{\rm DP} \rho)^2}{2} \ \frac{K_2(\alpha_{\rm DP} x)}
{1+\displaystyle \frac{(\alpha_{\rm DP} \rho)^2}{2}K_2(\alpha_{\rm DP} x)} \ ,
\eeq
that behaves as $K_2(\alpha_{\rm DP} |x|)$ at $x \gg \rho$~\footnote{In ref.
\cite{Dorokhov:1999ig}, in the context  of a constrained instanton
model, the authors propose Ans\"atze to account for
large-scale vacuum field fluctuations also by matching similarly large and 
short distances limits of their constrained instanton equation.
They also obtain  solutions decaying exponentially at large
distances.} and at short distances as:

\beq\label{BPSTlimit}
\phi\left(\frac{|x|}{\rho}\right) = \frac{\rho^2}{\rho^2+x^2}+O(\alpha_{\rm DP}x) \ ,
\eeq
where for $\alpha_{\rm DP}\, x \ll 1$ we indeed recover the BPST instanton. 
Remember that the zero $\alpha_{\rm DP}$ 
limit is also, from eq.~(\ref{baddens}), the zero density limit. 
Notice that the imposed constraint, expressed by eq.~(\ref{BPSTlimit}), of a  
behaviour {\it a la BPST} at $\alpha_{\rm DP}\, x \ll 1$
 is {\it de facto} the  scale invariance breaking condition in our approach. 
 It differs slightly from DP's proposal  $\phi(1)=1/2$
since $\phi(1) = 1/2-(\alpha_{\rm DP}\rho)^2/16+O((\alpha_{\rm DP}\rho)4)$. 
For the values of $\alpha_{\rm DP}$ and ${\overline \rho}$ which we will use to fit
the lattice data both are in practice equivalent. We show in fig. \ref{Fig0} a plot 
of this profile for $x/\rho>1/2$ compared to a numerical solution of Eq. (\ref{phi})
equal to $1/2$ at $x/\rho=1$ and going to $0$ at infinity (for
$\alpha_{\rm DP} \rho =0.675$). In the following, we will consider \eq{para} as our optimal
choice for the profile function.

\vspace*{0.4cm}

\section{Green functions}
\label{GFsection}

The goal of this section is to apply the parameterization in \eq{para} to fit our 
numerical results, obtaining thus $\rho$, $\alpha_{\rm DP}$ and the instanton density.
We start by writing the gauge field as the addition of a classical part,
$B_\mu^{a}({\bf x})$, and a quantum one, $Q_B({\bf x})$, depending
in general on the classical background:
\beq
A_\mu^a({\bf x})=B_\mu^{a}({\bf x})+ {(Q_B)}_\mu^a({\bf x}) \ ,
\eeq
then, for the two-point gluon Green function, we can 
write~\footnote{Crossed terms vanish if the background 
corresponds to a local minimum of the action i.e. to a classical
solution.}:
\beq
\langle A_\mu^a({\bf 0}) A_\nu^b({\bf x}) \rangle  = 
\langle B_\mu^{a}({\bf 0}) B_\nu^{b}({\bf x}) \rangle \ + 
\langle (Q_B)_\mu^{a}({\bf 0}) (Q_B)_\nu^{b}({\bf x}) \rangle \  ,
\label{propx}
\eeq
or after Fourier transformation, 
\beq
G^{(2)}_{\rm lattice}(k^2)=G^{(2)}_{(I)}(k^2)\ +\ G^{(2)}_{Q}(k^2)\ .
\eeq

The working hypothesis we derive from the interpretation of 
fig. \ref{Fig1}, is that in a certain region of momenta, say below 
$\sim$ 1 GeV, quantum effects are strongly suppressed and only 
classical properties are seen. Of course, we introduce this 
hypothesis based on a phenomenological observation, but it is 
not unconceivable that an intrinsically non-perturbative 
phenomenon like confinement could cause the disappearance 
of quantum correlations at distances larger than the confining scale. 

\subsection{Estimated corrections to the instanton liquid model}
\label{corr}
From our data we have been led to fit the bare Green functions 
to the r.h.s of eq. (\ref{propins}) in a range 
$k_{\rm min}< k<k_{\rm max}$ without any correction. 
It is far from obvious that we can neglect either quantum corrections
or lattice truncations of the quasi-classical model.  This
subsection is devoted to justify our choice. For the sake of clarity, 
we will pay the price of 
anticipating on the results of some fits which will be later detailed. 

\begin{figure}[ht!]
\begin{tabular}{c c}
\includegraphics[width=16pc]{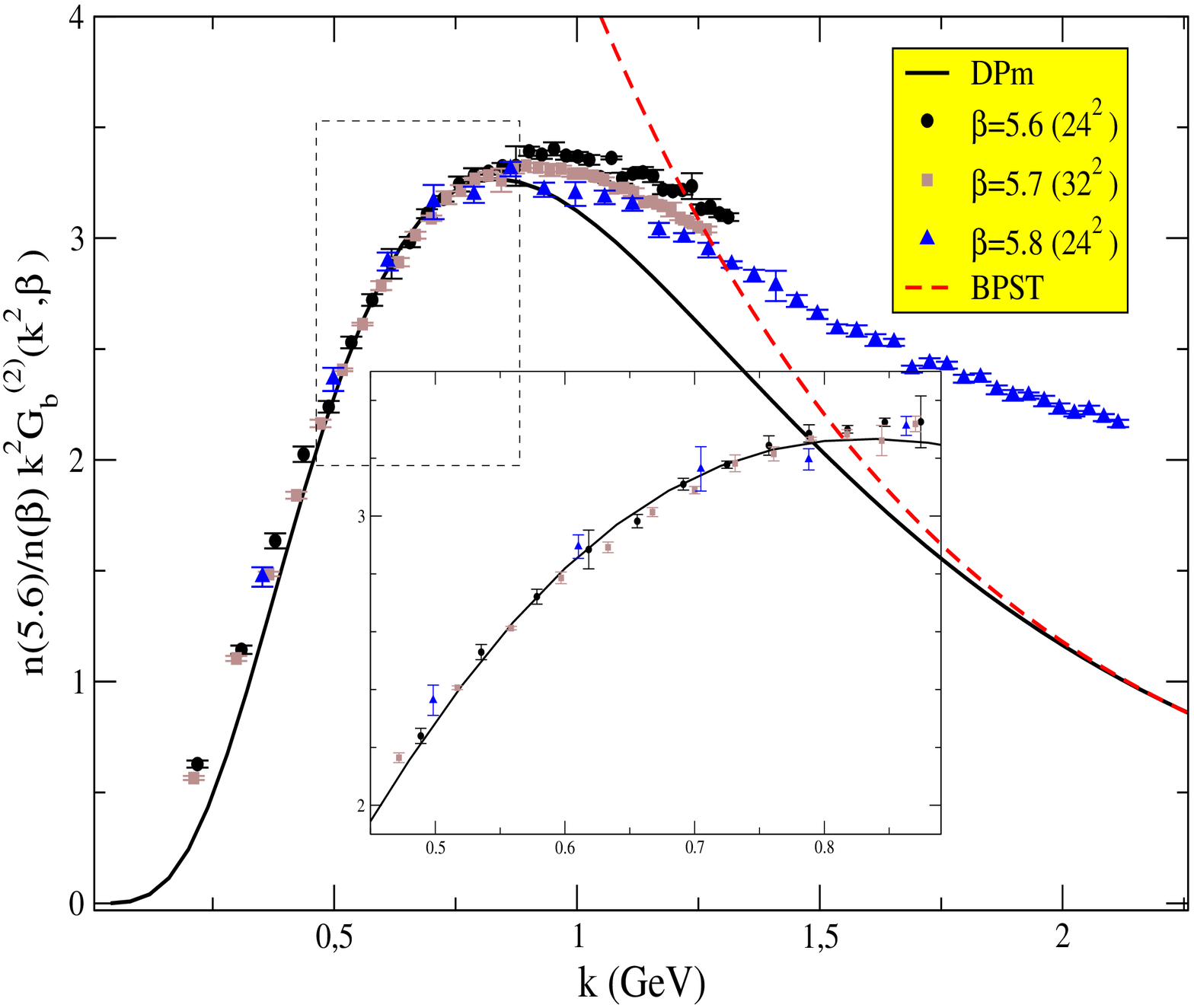}  & 
\includegraphics[width=16pc]{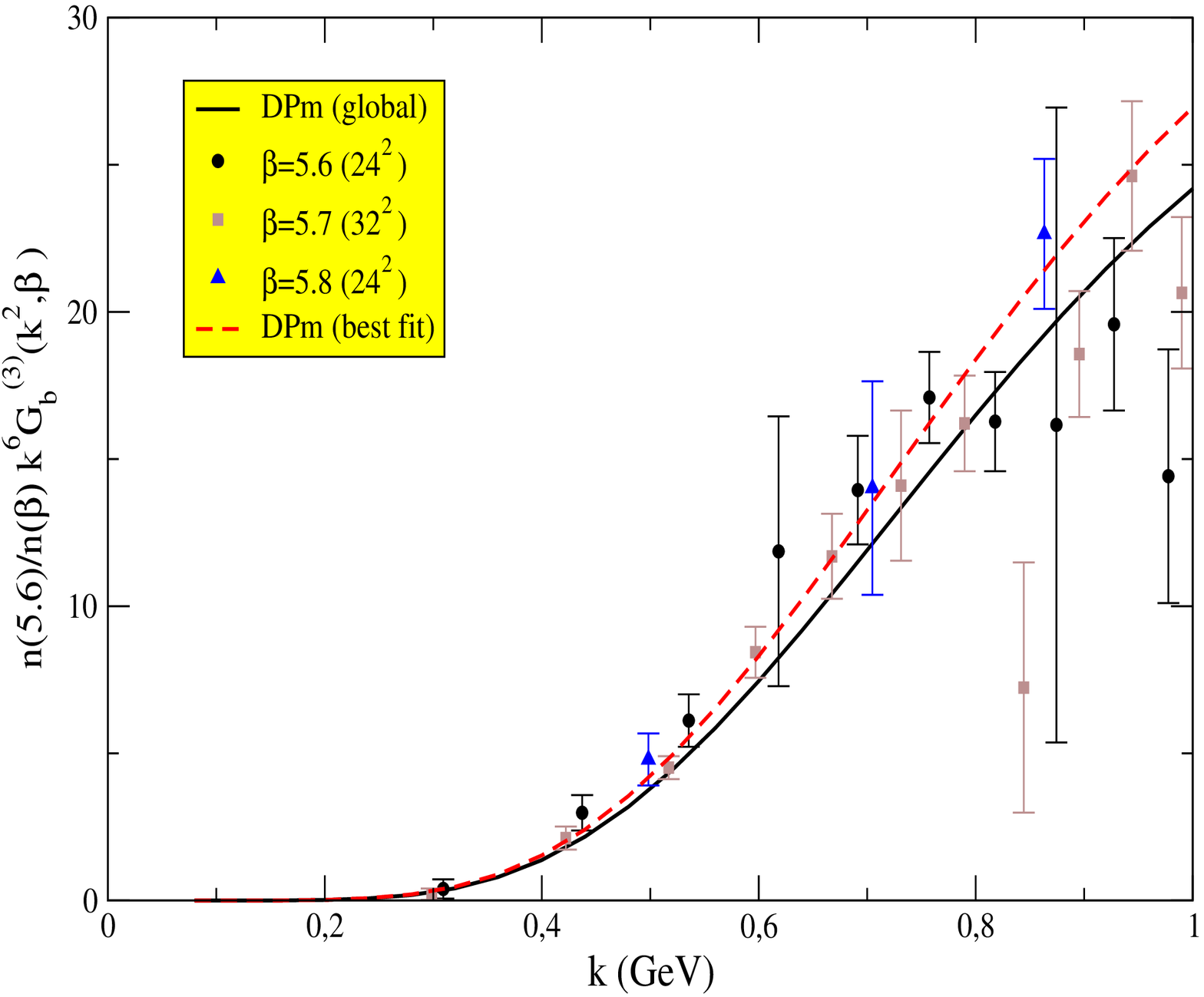} \\ 
(a) & (b) \\
\end{tabular}
\caption{\small \it (a) We plot the gluon propagator, $k^2G^{(2)}$,
for the three lattices after  a rescaling that matches data from
$\beta=5.7$ and $\beta=5.8$ to those from  $\beta=5.6$.  The dotted
line is obtained from the
BPST profile with $\rho=0.3$ fm  and the solid line is the best fit
using  our profile parameterization for the gluon
propagator. BPST cannot manifestly describe
lattice data. (b) The points are lattice evaluations of symmetric
vertex, $k^6G^{(3)}$. The solid line is the  best fit for the vertex
with our profile parameterization, the density being required  to be the
same for both propagator and vertex, and the dotted is the best fit
with a free density in the fitting.}
\label{fits}
\end{figure}

\begin{itemize}

\item[(i)] The Green functions $G^{(n)}$ estimated from three 
different lattice spacings match reasonably to each other, after
 a mere rescaling  to a common value for some $k_{\rm max}$ of
the order of 1 GeV (see Fig. \ref{fits}). These matching
coefficients are close to 1,  approximately in the ratios 0.95 /
1.0/ 1.05 for $\beta = 5.6, 5.7, 5.8$. 
 \item[(ii)] It is difficult to know precisely
 which mechanism drives these matching coefficients slightly away from 1:
 quantum corrections, ultraviolet/infrared cutoffs on quasi-classical 
 solutions. It is also impossible to know if these corrections are
 multiplicative, additive or of a more complex nature. 
 The main lesson is that these corrections are small,
 and we decide for convenience to describe them by 
 a multiplicative rescaling factor. 
  \item[(iii)]  This factor is defined  as      
  \beq\label{rescaling}
\Gamma^{(m)}(k^2,a^{-1}) \  \ = \  \
\frac{G^{(m)}_{\rm lattice}(k^2,a^{-1})}{G^{(m)}_{(I)}(k^2)} \ ,
\eeq
where $a^{-1}$ is the  regularization scale i.e. the inverse
lattice spacing  and $G^{(m)}_{(I)}(k^2), m=2,3$ are given in
eq. (\ref{propins}). The functions $\Gamma^{(m)}(k^2,a^{-1})\, n\,
\beta/6$~\footnote{The factor $\beta/6 \equiv 1/g^2$ comes from the
factor $g^2$ in \eq{propins}.},
$n$ being the instanton density, are plotted  up to an unknown
global constant in fig. \ref{ratios}(a).  $G^{(2)}_{(I)}(k^2)$ is
here the best fit to be discussed later. These plots show the good
matching of different  lattice spacings after performing 
a constant multiplicative rescaling (fig \ref{ratios}(b)
shows the result of this rescaling). They show  a wide
parabola having its minimum around  $k=0.65$ GeV, 
this minimum is around 7 to 8 which corresponds 
to the order of magnitude of the instanton density we will derive
(the other factors being close to 1). 

One possible
explanation of this parabolic behaviour is an expected increase of
quantum corrections towards larger $k$ and, towards lower $k$, a
relative increase of the ratio due to the fast decrease of the
denominator in \eq{rescaling}: for example the fast increase
 at low $k$  might be due to {\it additive} quantum corrections 
 which are only visible when the
quasi-classical background is very small. Additive corrections are
 anyhow necessary at $k=0$ since the lattice data are non vanishing
 while the denominator of  \eq{rescaling} is zero. Whether this non
  vanishing of the lattice propagator at $k=0$ is a finite volume 
  effect violating  Zwanziger's theorem~\cite{Zwanziger:gz}, 
  will not be discussed in this paper since, as already mentioned,
   we restrain from discussing the finite-volume sensitive region 
  below $\sim 0.4$ GeV.  

\vskip 1 cm
\begin{figure}[h]
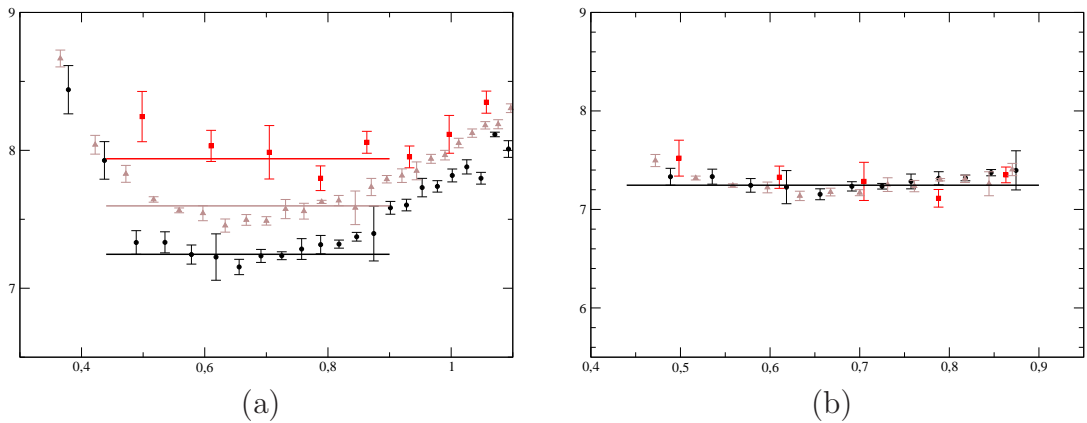

\begin{tabular}{c c c}
\includegraphics[width=16pc]{ratio.eps}  & & 
\includegraphics[width=16pc]{ratio2.eps} \\ 
(a) & & (b) \\
\end{tabular}
\caption{\small \it (a) We plot the ratios
$n\,\Gamma^{(2)}(k^2,a^{-2})\beta/6$ where $n$ is the instanton 
density and $\Gamma^{(2)}$ defined in eq. (\ref{rescaling}), for 
the three lattices, the  squares, triangles and circles
corresponding respectiveley to  $\beta=5.8$, $\beta=5.7$ and
$\beta=5.6$.  The horizontal axis represents the momentum in GeV.
 (b) The same points as in (a) but with the
multiplicative rescaling performed. Different $\beta's$ coincide
quite well, all show a parabolic type dependence in $k$.}
\label{ratios}
\end{figure}

\item[(iv)] If we stick to the momentum  range plotted in  fig.
 \ref{ratios}(b)  these ratios do not deviate from a constant by 
 more than a few percent. We will therefore perform our fits only
 in this  range $k_{\rm min} - k_{\rm max}$~\footnote{To be
 precise our fits  use the window $k_{\rm min}=0.44$ and $k_{\rm
 max}=0.89$.} and approximate  $\Gamma^{(2)}(k^2,a^{-1})$ by a
 constant:
 \beq
\label{hypo}
\Gamma^{(2)}(k^2,a^{-1}) \begin{array}{c}
  \rule[0cm]{0cm}{0.8cm} \sim \rule[0cm]{0.5cm}{0cm} \\
   k_{\rm min}^2\lwrsim k^2 \lwrsim
  k_{\rm max}^2 \end{array} \Gamma^{(2)}(a^{-1}) \ ,
\eeq
\item[(v)] Since from fig. \ref{ratios}(a) 
the different $\Gamma^{(2)}(a^{-1})$
differ also by no more than a few percent
 we decide to take from now on all these $\Gamma^{(2)}(a^{-1})$'s
 as equal at the cost of an expected discrepancy between the fitted 
 densities of a few percent. We take
\beq
\label{hypo2}
        \forall a^{-1},\quad \Gamma^{(2)}(a^{-1}) \ \ = 
         \ \Gamma^{(2)}\left(a^{-1}(5.6)\right)\ .
\eeq
where $a^{-1}(5.6)$ is the inverse lattice spacing for $\beta=5.6$.
\item[(vi)] We now wonder if the hypothesis eq.(\ref{hypo2}) can be
extended to  the coefficients $\Gamma^{(3)}(a^{-1})$. The small
dependence on $k$ and on $a^{-1}$ can be seen, although with larger
errors, in fig. \ref{fits}(b) where  the different lattice data seem
to fit  one common solid curve representing  the model. The next
question is the validity of the  hypothesis 
\beq
\label{hypo3}
 \Gamma^{(3)}(a^{-1}) \ \ = \Gamma^{(2)}(a^{-1}).
\eeq
To test the hypothesis $\Gamma^{(3)} \sim \Gamma^{(2)}$ we can look at 
fig.~\ref{Fig2}. The vertical axis corresponds to
the ratio $\Gamma^{(3)}(a^{-1}(5.6)/\Gamma^{(2)}(a^{-1}(5.6))$
or equivalently to  $n^{(3)}/n^{(2)}$ where we define $n^{(m)} \equiv n
\Gamma^{(m)}(a^{-1}(5.6))$, $n$ being the instanton density. The
horizontal axis corresponds to  the instanton radius $\rho$, supposed
to be the same for all instantons.  For that instanton radius  a best
fit of  the bare Green functions is performed according to the
formula:
\beq
G^{(m)}_{\rm lattice}(k^2,a^{-1})=G^{(m)}_{(I)}(k^2) \ 
\Gamma^{(m)}(a^{-1}(5.6))
\eeq 
with $G^{(m)}_{(I)}(k^2)$ given by eq. (\ref{propins}). The
plot~\ref{Fig2} shows for each $\rho$ an agreement between different 
$\beta$'s which is a surprise: it tells that the ratio
$\Gamma^{(3)}(a^{-1})/\Gamma^{(2)}(a^{-1})$ is almost independent of
the lattice spacing even though it does strongly depend on $\rho$. The
dotted curve shows the $\chi^2$ of the common fit  to $G^{(2)}$ and
$G^{(3)}$. {\it The smallest $\chi^2$ corresponds  to a ratio
$\Gamma^{(3)}(a^{-1})/\Gamma^{(2)}(a^{-1})$ ranging from 0.8 to 0.9, 
i.e. close to 1. This also
corresponds to a value of the radius around 0.3 fm which is the one
favored by phenomenology}. This leads us to consider 
eq.~(\ref{hypo3}) as quite reasonable.
 
 \item[(vii)]
 
 All the arguments up to now have shown an approximate equality to a
few percent of all $\Gamma^{(m)}(a^{-1})$'s. We are left with an
unknown global constant. The fact that these 
factors are so close to each other suggests that the corrections to the instantonic
contribution (whatever their origin may be) are small.
It is then natural to expect that these correction factors  are also close to 1.
 This is
 confirmed by a result which will be detailed at the end of this 
 paper via the following argument:  if
we initially take $\Gamma^{(2,3)} \neq 1$, apply MOM renormalisation
and then 
\eq{mom}, we obtain:
\beq\label{alphaLatt}
\alpha_{\rm Latt}(k) \ = \ \frac{k^6}{4 \pi} \frac{\left(
G_{\rm Latt}^{(3)}(k^2,a_{\rm Latt}^{-1})\right)^2}
{\left(G_{\rm Latt}^{(2)}(k^2,a_{\rm Latt}^{-1})\right)^3} \ = 
\underbrace{\frac{\left(\Gamma^{(3)}(a_{\rm Latt}^{-1}) \right)^2}
{\left(\Gamma^{(2)}(a_{\rm Latt}^{-1}) \right)^3}} \ \ \alpha_{(I)}(k) 
\eeq
where the term above the curly bracket in the r.h.s. introduces a
supposed-to-be-sub-leading  correction, depending on the
regularization parameter, that we neglect in our  analysis for the
present work (and did the same in~\cite{Boucaud:2002fx}).  Let us
anticipate some of our next results:  the reasonable agreement between the instanton
density obtained from the  coupling constant $\alpha_{\rm Latt}(k)$
in table~\ref{Tab2}  with the one in table~\ref{tabfits} implies a
ratio 
$(\Gamma^{(3)})^2/(\Gamma^{(2)})^3$ close to 1. The latter, combined with 
$\Gamma^{(2)}/\Gamma^{(3)}\sim 1$ presented in the preceding item,
 ends up with an approximate justification of our hypothesis
 that {\it all $\Gamma^{(n)}(a^{-1}), n=2,3\sim 1$ for all $\beta$'s
 in the range considered here}. 

\end{itemize}

Thus, our final working hypothesis is:
 \beq\label{HYPO} 
  \Gamma^{(m)}(k^2,a^{-1}) = 1 \quad {\rm for\ \  m=2,3\ \ and}\quad 
  k_{\rm min} < k < k_{\rm max}.
 \eeq
 with $k_{\rm min}=0.44$ GeV and $k_{\rm max}=0.89$  GeV.

\begin{figure}[h!]
\begin{center}
\includegraphics[width=20pc]{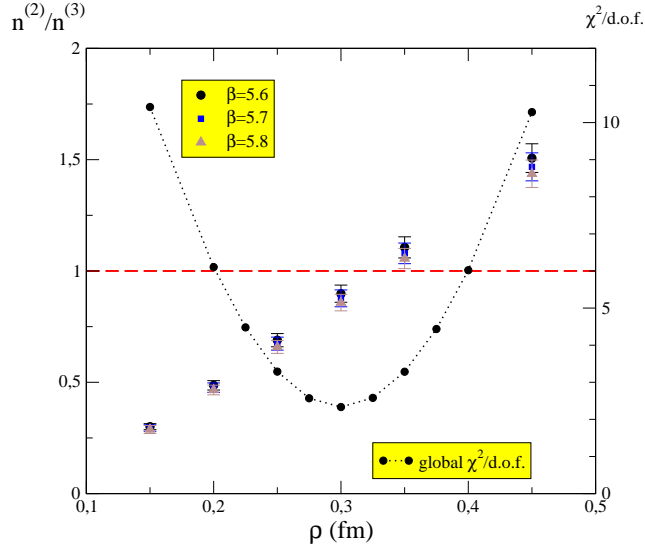}  
\caption{\small \it We plot the ratios $n^{(3)}/n^{(2)}$ where we
define $n^{(m)} \equiv n \Gamma^{(m)}(a^{-1}(5.6)$, $n$ being the
instanton density. These are  fitted from two-point and
three-point Green functions  for a fixed value of instanton radius,
$\rho$, represented  on the horizontal axis. The dotted line joins
the optimal $\chi^2/d.o.f.$  (on left  y-axis) as a function of the
radius.}
\label{Fig2}
\end{center}
\end{figure}

\subsection{Numerical results.\\}

The numerical data that we shall exploit systematicaly all over the paper result
from  two simulations on a $24^4$ lattice with bare coupling
constants given by  $\beta=5.6$ and $\beta=5.8$, and a simulation on
a $32^4$ lattice for  $\beta=5.7$\footnote{With the two lattices for
$\beta=5.6,5.7$ we simulate  practically the same physical volume and
are thus in position to check lattice spacing artifacts.}, in
the Landau gauge. The gluon correlations functions are obtained from only 20 field configurations, that 
are enough to manifest the classical background effects we search for. However, more field configurations 
are available for some of the lattices and we used them to check that those exploited are not particularly 
biased\footnote{Our Jacknife errors estimate also the dependence on the set of configurations.}.
We calibrate all  these simulations with the ratios
of lattice spacings for different $\beta$'s  given in ref.
\cite{Guagnelli:2000jw} and $a_{\rm Latt}^{-1}(\beta=6.0)=1.97$  GeV
(this last value was used in ref. \cite{Becirevic:1999uc} consistently
with the very precise measurement of the lattice spacing
resulting from a non-perturbatively improved action in ref.~\cite{Becirevic:1998jp}).

Coefficients in \cite{Guagnelli:2000jw} are fitted for
$\beta=6/g^2\ge 6$. In this work, we make use of rather low values
$\beta < 6.0$, in order to have larger volumes. These low values of
$\beta$ show for the quantity $\alpha_s$  a good  scaling with
results at  $\beta\geq 6.0$, a signal that these lower $\beta$'s provide
reasonable results. The risk with these simulations is the
extrapolation needed to calibrate the lattice  spacing, that might be
a non-negligible source of systematic error in our measures.

\begin{table}[ht]
\begin{center}
\begin{tabular}{||c||c|c|c|c|c||}
\hline
\hline
  $\rho$ (fm) & 0.2 & 0.25 & \bf 0.3 & 0.35 & 0.4 \\
\hline
\begin{tabular}{c|c} \rule[0cm]{1.2cm}{0cm} & $\beta=5.6$ \\ 
\end{tabular} & 26.5(1) & 13.18(5) & \bf 7.75(3) & 5.12(2) & 3.63(1) \\
\cline{2-6} 
\begin{tabular}{c|c} \cline{2-2} \rule[0cm]{0.3cm}{0cm} n \rule[0cm]{0.42cm}{0cm} & $\beta=5.7$  \\
\end{tabular} & 27.2(1) & 13.54(5) & \bf 7.98(3) & 5.28(3) & 3.76(2) \\
\cline{2-6}   
\begin{tabular}{c|c} \cline{2-2} (fm$^{-4}$) & $\beta=5.8$ 
\end{tabular} & 27.8(2) & 13.9(1) & \bf 8.20(7) & 5.41(4) & 3.86(3) \\ 
\hline
$\alpha_{\rm DP} \rho$ & 0.393(2) & 0.527(2) & \bf 0.675(2) & 0.836(3) & 1.005(3) \\
\hline 
$\chi^2/d.o.f.$ & 6.1 & 3.3 & \bf 2.3 & 3.3 & 6.0 \\
\hline
\hline
\end{tabular}
\caption{\small Best-fit parameters for several fixed instantons
  radii. The two- and three-point Green functions are fitted simultaneously 
  with the some instanton density. Note that the stated errors are only 
  statistical. }
\label{tabfits}
\end{center}
\end{table}

We collect the results of our fits for several fixed instanton radii
in  tab. \ref{tabfits}. An instanton density, although the same for
both Green  functions~\footnote{Notice the difference with
the fits presented in fig. \ref{Fig2} where different 
densities are assumed for $G^{(2)}$ and $G^{(3)}$.}, is independently fitted for each particular
lattice spacing. We  look thus for a remnant of a sub-leading
dependence on the regularization  parameter. The careful reader may
have noticed that the relative  density splitting between different
$\beta$'s is slightly smaller  (3 \% for $\delta \beta =0.1$) than
the splitting between the  linear fits in fig. \ref{ratios} (4.5 \%).
This is due the factor $g^2$ in eq. (\ref{propins}) (about 1.5 \%).

\paragraph{Crudeness of our quasi-classical approximation.\\}

It is worth to insist that our fit rely on the crude
approximation given by eq. (\ref{HYPO}) which is not valid beyond a few
percents as discussed at length in subsection \ref{corr}. This is the
origin of the difference  between the densities in tab.
\ref{tabfits}. This is also  the reason why  the minimal
$\chi^2/d.o.f.$ in table \ref{tabfits} is  of the order of  two
because we  force the overall factor  to be the same in fits for both
two- and three-point  Green functions. This tells about the crudeness
of completely neglecting  the sub-leading  contributions from quantum
fluctuations and lattice truncation of the classical solutions.  Much
better matchings might have been obtained had we relaxed the 
constraint eq. (\ref{HYPO}). We refrained from doing so because it
would  have needed additional input about quantum fluctuations and
lattice truncation of classical solutions which would have been mere
guesses and would not have yielded any stronger evidence.
 
At the present stage of the work, the best we can do is to assume 
eq. (\ref{HYPO}) and get what we believe to be a fairly coherent
description of our whole set of lattice data. The best-fit 
parameterization in Tab. \ref{tabfits} is the best we can achieve,
and it is not that bad,  although we know that the  density 
obtained there is affected roughly by a 20\% of systematic 
uncertainty\footnote{This uncertainty is estimated from 
the $n^{(3)}/n^{(2)}$ ratio at the minimum $\chi^2$ 
 in fig. \ref{Fig2}, ranging around 0.8--0.9 instead of 1.}. 

\subsection{QCD versus pure gluon-dynamics}

Our measured densities in tab. \ref{Tab1}, while in the
ballpark  of lattice estimations, give a contribution to the gluon condensate
$\VEV{g^2 {G_{\mu\nu}^{a}}^2} \sim 4 \Gev^4 \sim (1400 \ \Mev)^4$ 
for $n \sim 8 \,{\rm fm}^{-4}$, 
significantly larger than other estimates, based on QCD  sum
rules~\cite{Shifman:bx}, $\VEV{g^2 {G_{\mu\nu}^{a}}^2} \sim 0.5 \ \Gev^4 \sim
(840  \Mev)^4$, or more recently \cite{Narison:1996sg}, $\VEV{g^2
{G_{\mu\nu}^{a}}^2}  \sim 0.9 \Gev^4 \sim (970 \Mev)^4$. A significant difference between the 
pure Yang-Mills condensate and the one from QCD sum rules is however expected: in ref. 
\cite{Novi81} (see eq. (106)~) it is argued that light dynamical quarks might reduce the $G^2$ 
condensate by a factor 2 to 3. In fact, it can be argued that the low eigenvalues 
of the fermion determinant reduce the instanton contribution from the action and hence the 
resulting instanton density. In addition, some re-shaping of instanton profiles for a 
strongly correlated 
vacuum populated by fermions cannot be discarded. As a result, estimates in pure
gluon-dynamics should 
differ substantially from those in QCD and only the analysis of unquenched ($N_f \neq 0$)
lattices simulations of 
gauge fields could help us to quantify this discrepancy. 

Not only the same analysis of this paper for unquenched simulations could
quantify the discrepancy, but  we may also appeal to the connection of
$A^2$-condensate and the ILM approach:  the gluon condensate of dimension two
can be computed by extracting the momentum power OPE correction to  the
perturbative formula of Landau-gauge Green functions from their lattice
estimates above  $\sim 2.6$ GeV (see discusion in section
\ref{4-regimes})~\cite{Boucaud:2000nd}. This condensate is  to be computed at a
given renormalization point, $\mu_0$, lying on the momentun range of 
OPE-dominance. 

Now, if we run the renormalization point $\mu_0$ of the $A^2$-condensate down
to some instanton  scale, $\mu_I$, with the help of the one-loop renormalisation
goup equations (RGE), we obtain~\cite{Boucaud:2002nc}:

\beq\label{Ainst}
\langle A^2_{\rm inst} \rangle \equiv 12 \pi \rho^2 n \simeq \langle A^2_{R,\mu_I} \rangle 
\simeq \langle A^2_{R,\mu_0} \rangle \left( 1 + \frac {35}{44} \ 
\frac{\ln{(\mu_I/\mu_0)}}{\ln{(\mu_0/\Lambda_{\rm QCD})}} \right) \ ;
\eeq
where we match the OPE+RGE estimate to an ILM semiclassical one. Of course, the
latter is  a purely semiclassical estimate deprived of the UV-fluctuations
around the classical  minima which we assume to be subleading at the
appropriated instanton scales (see again  the discussion in section
\ref{4-regimes}). Thus, the instanton density can be  roughly estimated from
the dimension-two gluon-condensate. 

Although the gluon condensate  for quenched simulations has been largely
studied~\cite{Boucaud:2000nd} in literature,  unfortunatly only a very preliminar 
analysis of the MOM QCD coupling constant from a lattice simulation  with two
dynamical-quark flavours is available~\cite{Unquenched}.
Furthermore, this analysis is performed for such high quark masses that
no effect of instanton  suppression is to be expected and indeed no
significant effect has yet been seen. Progress in the unquenched determination 
of the gluon condensate will be, of course, welcome.

Moreover, the QCD sum rules are not so accurate and neither are 
our density estimates which, for instance, would be modified by the presence of 
other ``instanton-like structures'' or instantons deformations.

{\it Altogether our density estimate is close to the maximum acceptable: with a
density of 8 per fm$^4$  the average distance between two neighbouring
instanton centers is of the order of 0.6 fm i.e. twice  the average radius, a
really dense packing. }

\section{The effect of the instanton radius dispersion}\label{radii}

We have performed satisfactory and independent fits for the
three  different lattices from $\alpha_s$ to the $k^4$-formula, eqs.
(\ref{alpha}, \ref{delta}), and obtain the estimates for the
instanton  density shown in the second column of Tab. \ref{Tab1}. 
They differ by  about a 30
\% from the profile dependent ones obtained from 
$G^{(m)}$ and presented in  table \ref{tabfits}. The fit
of $\alpha_s$ is done on a rather firm ground since, contrarily to
the fits of $G^{(m)}$, the prediction  does not at all depend on the
profile neither on the radius but only on eq.(\ref{delta}), of course on
the hypothesis  that the radius distribution  is a delta function
{\it i.e.} that all the radii are equal.  $\alpha_s$ is therefore the
best quantity to estimate the corrections to this rather drastic
hypothesis.  If the $k^4$ law does not depend on the instanton
profiles but only on the $\delta$-function distribution, the corrections 
do~\footnote{Obviously the influence of the radius distribution
$\mu(\rho)$ will not be independent of the profile (Remember
(\ref{propins})).}. 

\begin{figure}
\begin{center}
\begin{tabular}{c}
\includegraphics[width=25pc]{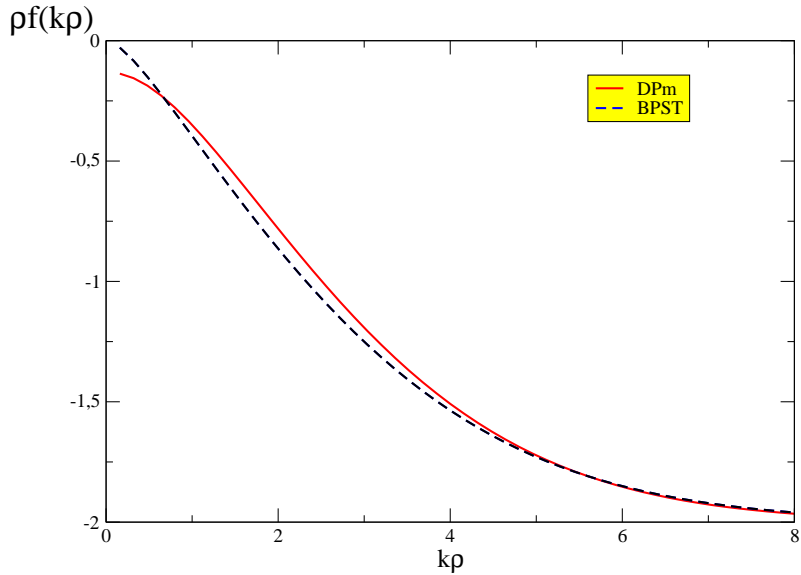} 
\end{tabular}
\caption{\small{\it The function $\rhob f(k \rhob)$ defined in the text in terms of $k\rhob$ for 
both DP-inspired and BPST profiles.}} 
\label{Xplot}
\end{center}
\end{figure}

First, we face the problem of computing the corrections to the previous 
$\delta$-function result from a more general optics: we just consider 
some small width for the distribution, $\delta \rho$, and compute in 
perturbations of $\delta \rho/\rhob$ around the average radius, 
$\rhob$. We can write:

\beq
< \rho^{3m} I^m(k \rho) > \ = \ I^m(k \rhob) \ < \rho^{3m} > \ + \ 
\frac d {d\rho}I^m(k \rhob) \ < \rho^{3m} (\rho - \rhob) > \nonumber \\
+ \ \frac 1 2 \frac {d^2}{d\rho^2} I^m(k \rhob) \ < \rho^{3m} (\rho - \rhob)^2 > \ + \dots \ .
\eeq
we take $\rho = \rhob + \delta\rho$ , assume the distribution to be symmetric (at least for 
small perturbations) around the peak, {\it i.e.} $< \delta\rho >=0$ and then obtain:

\beq
< (\rhob + \delta\rho)^n > \ = \ \rhob^n \left( 1 + \frac {n(n-1)} 2 \ \frac
{\delta\rho^2} {\rhob^2} 
+ {\cal O}\left( \frac {\delta\rho^4}{\rhob^4} \right) \right) \ , \nonumber \\ 
< (\rhob + \delta\rho)^n \delta\rho > \ = \ n \rhob^{n+1} \frac{\delta\rho^2}{\rhob^2} \ + 
{\cal O}\left( \frac {\delta\rho^4}{\rhob^4} \right) \ , \nonumber \\
< (\rhob + \delta\rho )^n \delta\rho^2 > \ = \ \rhob^{n+2} \ \frac{\delta\rho^2}{\rhob^2} \ + 
{\cal O}\left( \frac {\delta\rho^4}{\rhob^4} \right) \ .
\eeq
Thus, we derive for the QCD MOM coupling constant defined in \eq{mom} the following expression:

\beq
\alpha(k) \ = \ \frac {k^4}{18 \pi n} \ \left( 1 \ + \ \frac {\delta \rho^2}{\rhob^2} \ 
\left( 12 + 12 \rhob f(k \rhob) + 3 \rhob^2 f^2(k \rhob) \right) + 
{\cal O}\left( \frac {\delta \rho^{4}}{\rhob^{4}}\right) \right) \ ,
\label{alpha-k4ex}
\eeq
where all the dependence on the profile function comes through the function $f$ 
defined as

\beq\label{f}
f(k \rhob) \ = \ \left. \frac d {d\rho} \ln \int_{0}^{\infty} z dz J_2(k \rho z) \phi(z) 
\ \ \right|_{\rho=\rhob} \ .
\eeq
In particular, for the BPST profile, we find:

\beq\label{fBPST}
\rhob f(k \rhob) \ = \ -2 \ \frac {1 - \displaystyle \frac {k^3\rhob^3} 8 
\left( K_1(k \rhob) + K_3(k \rhob) \right) } {1 - \displaystyle \frac {k^2 \rhob^2} 2 K_2(k \rhob)} \ .
\eeq
This function $\rhob f(k \rhob)$ takes only negatives values for all $k \rhob$. 
The same is obtained by applying the DP-inspired profile parametrization 
defined in \eq{para} (where we take $\alpha\rhob=0.68$ from the previous section) 
to \eq{f} and, as can be seen in plot \ref{Xplot}, the functions $f$ obtained from 
both profiles evolve rather close to each other for all $k \rhob$. 
This suggests a pattern that \eq{alpha-k4ex} obeys in rather general 
manner\footnote{see appendix \ref{A}.}: 
the $k^4$-power law is corrected by a positive 
$k \rhob$-independent term and by a negative one depending on $k \rhob$; the first one gives 
some pre-factor larger than 1 to the $k^4$ formula and the second can be numerically
simulated rather well over the window $(0.4-0.9)$ GeV by reducing\footnote{It is in fact 
a matter of power reduction because $k \rhob > 1$ in the most and more numerically relevant part of the 
fitting interval, for $\rhob=1.5$ GeV$^{-1}$.} the power on $k$,

\beq
\alpha(k) \ \simeq \ \frac c {18 \pi n \rhob^4} \left( k \rhob \right)^{4-\varepsilon} \  .
\label{alpha-k(4-e)} 
\eeq
This last result, \eq{alpha-k(4-e)}, altogether with the 
constraints $c>1$ and $\varepsilon > 0$, are the main and more general result of the section. 
The values of $\varepsilon$ and $c$ depend on the particular profile and on the width of the 
radii distribution, as given by eqs. (\ref{alpha-k4ex},\ref{f}) in the small width limit. 

We will now proceed as follows. We will first estimate the parameter $\varepsilon$ by fitting 
our lattice data for MOM QCD coupling constant to the formula \eq{alpha-k(4-e)}, 
where $n/c$ is to be taken as a free parameter to be tuned. Then, we will estimate the width 
and the instanton density, $n$, with the help of eqs. (\ref{alpha-k4ex},\ref{f}), {\it i.e.} in the  
small width approximation. Finally, the consequences of employing more realistic radii 
distributions will be preliminarily discussed, details being in appendix \ref{B}.

\paragraph{Lattice results\\}

In this subsection, we first collected in tab.
\ref{Tab1} the results for the  density obtained through the fit of 
the MOM QCD coupling constant defined in \eq{mom} to the $k^4$-power 
law in eqs. (\ref{alpha},\ref{delta}), for our three
lattice data set.

\begin{table}[h]
\begin{center}
\begin{tabular}{||c||c||}
\hline
\hline
  Lattice & n (fm$^{-4}$)  \\
 \hline
5.6($24^4$) & 5.2(6)  \\
\hline
5.7($32^4$) & 6.7(4) \\
\hline
5.8($32^4$) & 6(1) \\
\hline
\hline
\end{tabular}
\caption{\small \it Results for the density obtained from fitting lattice
coupling constant to  eqs. (\ref{alpha},\ref{delta}) i.e. assuming a  $k^4$
behavior. Errors quoted are only statistical and computed by the jackknife  method}
\label{Tab1}
\end{center}
\end{table}

\noindent Then, we include in tab. \ref{Tab2} the results for the best fits of
the same  QCD coupling constant lattice data to the power law formula 
of \eq{alpha-k(4-e)}.

\begin{table}[h]
\begin{center}
\begin{tabular}{||c||c|c|c||}
\hline
\hline
  Lattice & \rule[-0.4cm]{0cm}{1cm} $\displaystyle \frac n c$ (fm$^{-4}$) & 
$\varepsilon$ & $\chi^2$/d.o.f. \\
 \hline
5.6($24^4$) & 6(2) & 0.2(4) & 0.068   \\
\hline
5.7($32^4$) & 7(3) & 0.1(5) & 0.36  \\
\hline
5.8($24^4$) & 5(1) & 0.2(6)  &  0.007  \\
\hline
\hline
\end{tabular}
\caption{\small The same that in tab. \ref{Tab1} but here fitting to
\eq{alpha-k(4-e)} for $\rhob=1.5$ GeV$^{-1}$. The  third row
shows the $\chi^2$/d.o.f. The fitted powers are then 
lower than 4 but  the errors show that the power 4 is not excluded
 by our fits.}
\label{Tab2}
\end{center}
\end{table}

\noindent We use for both the same fitting window, $0.44 < k < 0.89$, used in section
\ref{GFsection}  to obtain the best profile parameterization. The $\chi^2$/d.o.f.'s in
tab. \ref{Tab2} are tiny. This may be due to: (i) the strong
correlation of our lattice estimates of the coupling for the  different
momenta, all computed from the same set of gauge field
configurations~\footnote{Still, our jackknife estimate of the statistical
error is reliable because, if the global average for the different momenta are
correlated, the averages for different sets of gauge configurations are not.};
(ii) the small number of points  in our fitting window (only 3 for
$5.8(24^4)$).

The best-fit parameters computed lattice by lattice in tab. \ref{Tab2}
manifest no appreciable  systematic deviation\footnote{As expected since we
found the sub-leading lattice-spacing dependent corrections to be small in
\eq{alphaLatt}.}. A global fit for the whole set  seems thus to
be appropriate. The resulting power deduced from a global fit is:
\beq\label{pow}
3.91 \pm 0.45 \ \ ,
\eeq
which is {\it compatible with both the general constraint $\varepsilon > 0$ and
also with a $k^4$-power law}.  The $\chi^2$/d.o.f. is 0.39 for this
global fit. In ref. \cite{Boucaud:2002fx}, the same global  fit
was performed using also a few other lattices (we added estimates of $\alpha_s$
over smaller volumes and larger $\beta$'s: 5.7(24$^4$), 
5.9(24$^4$), 6.0(24$^4$,16$^4$); these, having only one point
inside the fitting window,  have been discarded from the present work.
The result of the latter analysis was {\bf
3.82(8)}  for the fitted power. The error was there clearly
underestimated\footnote{We used the criterium of assuming one 
standard deviation as $\chi^2=\chi_{\rm min}^2+1$, 
which is biased by the strong correlations between data for different 
momenta as just discussed.} but the evaluations of the coupling constant from
larger  $\beta$'s do not practically affect the central value.
This fact indicates that exploiting  the small $\beta$'s used in the
present work does not seem to introduce any sizable bias, at least over 
the low-momentum region, on the determination of $\alpha_s$.

\paragraph{Instanton density in the small width approximation \\}

The global fit for the whole set of lattice data yields $\varepsilon \simeq 0.1$ 
for the central value estimate of that parameter. If we apply the exact result for 
$f$ in \eq{f} to \eq{alpha-k4ex} and use the latter to estimate $\delta\rho$ 
through its best numerical matching to the power formula \eq{alpha-k(4-e)} with 
$\varepsilon=0.1$, we will obtain:

\beq
\frac {\delta\rho^2}{\rhob^2} \simeq 0.01 \ .
\eeq
This results implies that

\beq
c \simeq  1 + 12 \ \frac {\delta\rho^2}{\rhob^2} \simeq 1.12 \ .
\eeq
Thus, densities obtained from tab. \ref{Tab2} will be about 10 \%  
larger than those in tab. \ref{Tab1} because of the 
factor $c$ in front of the power of $k$. If we compute numerically the function $f$ 
for the DP-inspired profile and follow the same procedure to estimate the width, the parameter 
$c$ and hence the corrections to the instanton density estimates from the $k^4$-law in 
tab. \ref{Tab1}, we will obtain roughly the same results. This could be expected after checking the 
rather similar behaviours of the function $f$ obtained from both profiles in plot \ref{Xplot}.
This means consequently that, once some small width distribution 
is assumed and practically not regarding to the profile, the estimated instanton 
density is $n\sim 6-8$ fm$^{-4}$.

All these estimates are anyhow in the right ballpark as far as different  
arguments~\cite{Negele:1998ev} seem to point
towards an instanton density of a few fm$^{-4}$~'s. 

\paragraph{Discussing about realistic distributions. \\}

After estimating the effects of a small width distribution, we will discuss 
the consequences of some instanton distributions proposed in the literature.

It is well known that the one-loop
tunneling amplitude of classical minima in gauge
theories gives the  standard growth $\mu(\rho) \sim \rho^6$ for
instanton radii  distribution~\cite{'tHooft:fv}. 
The classical  interaction of instantons in the background is thought to 
introduce  small-distance repulsion that strongly suppress large
instantons  and leads to a well-behaved partition
function~\cite{Schafer:1996wv}.
Applying the Feynman variational principle to a liquid of BPST instantons, 
Diakonov \& Petrov found 
indeed that this infrared growth is balanced as
follows~\cite{Diakonov:1983hh}:
\beq\label{distr}
\mu(\rho)=2 \frac{\left(\frac 2 7 \overline{\rho^2}\right)^{-7/2}}
{\gamma(7/2)} \ \rho^6 
\exp{\left(-\frac{7 \rho^2}{\rule[0cm]{0cm}{0.4cm} 2 
\overline{\rho^2}}\right)} \ ,
\eeq
where $\overline{\rho^2}=\int_0^\infty d\rho \rho^2 \mu(\rho)$ 
and $\gamma(n)=\int_0^\infty dt e^{-t} t^{n-1}$ is the standar Euler's
gamma function and 

\beq\label{rho2}
\left( \rhodb \right)^2 \ = \ \frac{7}{2n\gamma_0^2 \beta(\rhob)} \ .
\eeq
A very important remark is that eqs. (\ref{alDP}), particularized for BPST 
profile\footnote{In the 
BPST limit \eq{alDP} becomes \eq{baddens} where 
$\int dx^2 \phi^2=\rhodb$.}  and using $\beta(\rho)\simeq\beta(\rhob)$, 
combined with eq.~(\ref{rho2}), leads to avoid the dependence on 
the density and to derive a prediction 
for the previously fitted parameter $\alpha_{\rm DP} \rho$, 

\beq\label{good}
\alpha_{\rm DP} \sqrt{\rhodb} \ = \ \sqrt{\frac {14} {3 \beta(\rhob)}} \ \simeq 0.6 
\eeq
where we use $\beta(\rhob) \simeq 15$ as in \cite{Diakonov:1983hh}. The agreement of such 
a prediction with our lattice estimate is remarkably encouraging since, for
our non-BPST profile, this relation has been numerically checked and remain
approximately valid.  

The goal of this section is anyhow to analyze the impact of the corrections to the equal 
radii approximation that we use to describe the low momentum behaviour of our lattice Green 
functions. For this purpose, we remark that  
the distribution given by \eq{distr} is rather asymmetric around 
$\rhob$, which is in practice usually defined as $\sqrt{\rhodb}$.
Its width, measured as

\beq
\frac {\delta\rho^4} {\rhob^4} \ = \  
\frac {\overline{\rho^4} - \left( \rhodb \right)^2} 
{\left( \rhodb \right)^2} \ = \frac 2 7 \ , 
\eeq
leads us to conclude that the previous small approximation could have missed some not negligeable 
correction. This is why we compute in appendix \ref{B} the Green functions and the MOM 
QCD coupling constant on the background of an instanton liquid with radii distribution 
given by \eq{distr} and fit the latter to the power formula given by \eq{alpha-k(4-e)}. We do it 
for BPST and DP-inspired profiles and collect the results in tab. \ref{Tab-ce}. As can be learned 
from the table, the main conclusions of the small width analysis remain unchanged except for the 
fact that the estimate of the density increases up to $\sim 7-10$ fm$^{-4}$ as a consequence of 
the values of the parameter $c$.

\begin{table}[h]
\begin{center}
\begin{tabular}{||c||c|c||}
\hline
\hline
  Profile & $c$ & 
$\varepsilon$ \\
 \hline
DPm & 1.55 & 0.17   \\
\hline
BPST & 1.52 & 0.25  \\
\hline
\hline
\end{tabular}
\caption{\small Values of the parameters $c$ and $\varepsilon$ in \eq{alpha-k(4-e)} obtained in the 
appendix \ref{B} for our DP-inspired and BPST profiles.}
\label{Tab-ce}
\end{center}
\end{table}

Furthermore, the instanton radii distribution has to be varied altogehter with
the instanton profile 
and density in the Diakonov \& Petrov formalism. These authors found 
\cite{Diakonov:1983hh}: 

\beq\label{DPfre}
\mu(\rho) \ \propto \ \frac{\rho^{11\; s(\rho)}}{\rho^5} 
\exp\left( - \beta n < \Gamma(\rho) \rho^2 > \Gamma(\rho) \rho^2 \right) \ ,
\eeq
for the radii distribution of what they call ``fremons'', {\it i.e.} the pseudo-instantons obeying 
the profile equation in (\ref{DPeq}), where $\Gamma(\rho) \rho^2 = \gamma_0 \int dx^2 \phi^2(x/\rho)$. 
$s(\rho)$ is a function on $\rho$ introducing a very small collective correction to the standard 
power $\rho^6$.

Two aspects which have lead Verbaarschot~\cite{Verbaarschot:1991sq} 
and Shuryak~\cite{ILM,Shur-SU2} to criticise 
the DP formalism are visible in the ``fremons'' radii distribution in
\eq{DPfre}:
(i) the distribution concerns the radius, $\rho$, of one particular instanton immersed in some 
``bath'' of instantons with average radius, $\rhob$; how does one combine the two scales in the 
game, $\rho$ and $\rhob$, to define the instanton scale for $\beta$ ? 
Why, for instance, use $\beta(\rho)$ instead of $\beta((\rhob \rho)^{1/2})$ ?
(ii) Such an instanton scale is furthermore of the order of 0.3 fm and hence the coupling $g$ in $\beta$ 
is to be computed out of the perturbative regime!

Any detailed evaluation of the impact of the ``fremon'' distribution implies answering these 
questions, in particular knowing the behaviour on $\rho$ of $\beta(\rho)$ and how it modifies 
the gaussian in \eq{distr}. However, this is out of the scope of the present work. We adopt here a 
``{\it phenomenological}'' point of view: the DP formalism offers a large-distance damping for 
the profile that properly corrects the low momentum behaviour of Green functions and leads to 
a satisfactory description of our lattice data. The large-distance behaviour of the profile is 
governed by the parameter $\alpha_{\rm DP} \rho$ which is rather well estimated when we 
put $\beta(\rhob)\simeq 15$ in \eq{good} according to \cite{Diakonov:1983hh}.
 Moreover, had we assumed $\beta(\rho)\simeq \beta(\rhob)$ 
and $\Gamma(\rho)\simeq 1$ (this is strictely true for the BPST profile) we would recover the 
distribution in \eq{distr}.

As a final remark, if we try to compute the instanton density through any of
the two eqs.  (\ref{baddens},\ref{rho2}) by employing our best-fit parameters
in the previous section, we  will obtain $\sim 2$ fm$^{-4}$. Is something still
missing in the DP approach? As a matter of  fact, the third main aspect
criticised by the authors above 
mentioned~\cite{ILM,Shur-SU2,Verbaarschot:1991sq} is the too strong instanton
interaction strength resulting from the DP formalism and it should be  noted
that {\it a reduction of $\gamma_0^2$ by a factor three or four would allow all
the pieces of the puzzle to match}. Thus, our analysis detects somehow this
effect pointed out  in the above references regarding the instanton-antiinstanton
interaction.  It remains to investigate whether our results are compatible
with  those, for instance in ~\cite{Verbaarschot:1991sq}, where the
Stream-line approach is used  (the author found that $\gamma_0^2$ is about one
order of magnitude smaller than the one from DP).

We should recall however our ``phenomenological'' point of view: does the sum-ansatz of 
Diakonov \& Petrov, after profile variation, take into account the effects included in the 
Stream-line analysis, except for some effective rescaling of $\gamma_0$ ? 
A preliminar analysis of this question is avanced in appendix \ref{C}, where we show how one can 
build the gauge field through the Shuryak's ratio-ansatz~\cite{Shur-SU2}, 
that gives an approximative solution to the Stream-line 
equation, and match the sum-ansatz with our DP-inspired instanton profile in both near the center of 
any of the instantons and far from all them. This could explain why the independent-pseudoparticle
approach given by the sum-ansatz accounts for the low momentum behaviour of the MOM QCD coupling constant 
and Green functions. This point, of course, deserves further attention.

\paragraph{}
In summary, we dedicated this section to analyze the effects of radius dispersion on the 
$k^4$ power-law that the instanton approach predicts for the MOM QCD coupling constant.
Within the small width approximation we establish that the two main effects of radius dispersion 
appears in our fitting window as a reduction of the power of $k$ and as an overall pre-factor larger 
than 1. In particular, the pre-factor does not depend on the particular profile function and 
corrects the estimate of the instanton density from the MOM QCD coupling constant that becomes 
then compatible with those directly obtained from Green functions in the previous section, 
$\sim 8 $ fm$^{-4}$.
In fact, this instanton-density estimate is rather general and, as far as the low momentum behaviour 
of Green functions is dominated by the gauge field at large distances, is not only independent of 
the profile function but neither is affected by the sum-ansatz approach we used. Nothing seems to change 
by the use of realistic distributions. Futhermore, the analysis of radius dispersion in terms of 
realistic distributions gives us the possibility of connecting the ``dynamical'' parameter 
$\alpha_{\rm DP} \sqrt{\rhodb}$ with only the instanton coupling throug $\beta$ and leads to an 
estimate of $\simeq 0.6$ for the former, in remarkable agreement with our fits.

\section{Discussions and Conclusions}
In a previous paper~\cite{Boucaud:2002fx}, we had shown a $k^4$ dependence  of the coupling
constant in MOM scheme, which gave us a rather clean indication of the
dominance of semiclassical solutions over low energy QCD dynamics,  and some
indication that these solutions might be instantons. 
 
In order to go further in the understanding of this very remarkable feature, we
tried here a direct description of gluon Green functions in this framework.
In order to  achieve this, contrarily to what happens for the MOM coupling
constant, the knowledge of the  instanton profile is mandatory. The
single-instanton BPST profile clearly fails to describe  the low momentum
behavior since it predicts a divergent gluon
propagator for $k \to 0$ in full  contradiction with the lattice results.
This is understood as an effect of the deformation of instantons far from 
their centers due to the influence of other instantons. 
 Indeed, the BPST singularity of the gluon propagator is corrected when
effects of instanton interactions are taken into account through  a
parameterization of the profile derived from variational
methods~\cite{Diakonov:1983hh}, and it leads to a successful description
of the low momentum behavior of lattice two- and three-point
Green functions (see fig. \ref{fits}). 
At this point, we cannot exclude other "instanton-like structures" as, 
for example, a significant amount of merons~\cite{Callan:qs,Lenz:2003jp} which would of course 
modify our density estimates. 

We fit our lattice data for the Green functions in an instanton liquid
model (ILM), below energies of $\sim 0.9$ GeV, and above $\simeq 0.4$ GeV.
 Below $\simeq 0.4$ GeV we have too few lattice data and furthermore 
the lattice computed gluon propagator gives a non
null value for the $k\to 0$ limit, in contradiction with the expectation of an 
ILM:

\beq 
G^{(2)}_{(I)} \sim \left\{ \begin{array}{l} k^{-2} \quad \mbox{\rm for BPST} \\
k^2 \quad \mbox{\rm for DP} 
\end{array}
\right.
\eeq
Futhermore, the quantum corrections to the semiclassical solutions should be more visible
when the latter vanish~\footnote{In ref. \cite{Mygdal}, the authors discuss about the impact of 
quantum effects on the large distance regime of instanton solutions.}. In fact, a signal that instanton 
approach could be missing 
some mechanism in the very low momentum regime is the fact that a very recent analysis within the  
SD approach leads to\footnote{It is possible to find in the literature previous estimates for the critical 
exponenent, $\kappa$, under different approximations' schemas as, for instance, 
$\kappa \simeq 0.92$~\cite{vonSmekal:1997is,vonSmekal:1997is2}, 
$\kappa \simeq 0.77$~\cite{Atkinson:1997tu} or 
$\kappa=1$~\cite{Atkinson:1998zc}.}

\beq
G^{(2)}_{\rm SD} \sim \left( k^2 \right)^{2 \kappa -1} \ ,
\eeq
with $\kappa \simeq 0.595$~\cite{Lerche:2002ep}. An also recent SD-motivated work, based on rather 
general arguments, points towards $\kappa = 1/2$~\cite{Kondo:2003sw}.
Last but not least, this region is expected to
be strongly finite-volume sensitive and deserves a special study.

We have discussed the effect of the instanton radius dispersion. We find
 a subdominant but non
negligible influence of the radius dispersion on the coupling constant:  
We find that a coherent description of both two- and three-point Green functions only emerges 
when the instanton radius reaches the vicinity of 1/3 fm, that is the value derived from 
phenomenological arguments. Furthermore, after invoking a radius
distribution  obtained from DP's variational methods\cite{Diakonov:1983hh},
we correct the  previously predicted $k^4$-power followed by $\alpha_s(k)$ by a slight
reduction of the power ($\sim k^{3.83}$) and a 40 \% increase of the prefactor.  
The fits of all our lattice data to a free-power law on $k$ always produces a best-fit exponent
lower than 4 but statistically compatible with 4 (We obtain 3.91(45)
from a global fit combining all our lattice data sets). This seems to 
confirm the trend predicted from radius distribution, but the small number of
points in the region of interest and the big errors lead to a large 
uncertainty that prevents us from a more conclusive statement. Once this radius dispersion 
is considered we get a good agreement of the density derived directly from a fit of the 
two- and three-point Green functions and from a fit to $\alpha_s(k)$.

The lattice spacings used in this study have been chosen rather 
large since we wished to reach low momenta with a not too large number
of lattice points. It might be feared that these values 
are too far from the continuum limit, too much in the strong coupling regime, 
 to be reliable. We did not see any sign of a non-smooth dependence
   of any quantity on the lattice spacing. However, for safety, it is 
   advisable (and under progress) to follow-on these studies at, say,
 $\beta=6.0$ with as large a physical volume as used here. 
 Larger statistics and larger physical volumes at a given lattice spacing
 are also needed.

The lattice gluon propagator at $k\to 0$, has also  open
problems, for example in relation to the Zwanziger problem \cite{Zwanziger:gz}
and critical exponent in relation to instantons. Other  theoretical questions 
are pending such as the reason why quantum fluctuations only play a sub-leading
role in this momentum range~\footnote{If we naively add quantum and classical
contributions to the gluon propagator (Eq. (\ref{propx})), the perturbative
quantum contribution would diverge.}. 

\paragraph{}

To briefly conclude, ILM can describe the low-momentum behaviour of gluon 
Green functions after modelling the instanton profile within the DP approach. 
Three are the parameters playing the game: $\alpha$, the instanton radius and 
density. The first one can be satisfactorily computed within the DP framework, 
the phenomelogical estimate of instanton radius leads to the best fits to lattice 
data and only the fitted density is much higher than its phenomenological estimate 
(although in agreement with another lattice estimates). Such a high instanton density 
leads to a large instanton packing fraction that makes hard a simple semiclassical 
approach to the gluon dynamics. However, instanton density should be reduced by 
including light dynamical quarks and it can not be then excluded that the semiclassical 
mechanism we use here might give account of the phenomenological value. Our 
large packing fraction do not rule out a semiclassical approach to the full-QCD 
dynamics.

Still we believe that we have given
a series of rather convincing evidences  of the influence
of instanton-like structures  on the low-energy QCD 
and more precisely on the low-momentum behavior of gluon Green
functions.

\appendix

\section{Small and large momentum regimes of $f$}
\label{A}

The goal of this appendix is to perform the analysis, based on rather general grounds, of the 
small and large momentum regimes of the function $f$ 
defined above in \eq{f}. We will assume both regimes to be controlled by the
Bessel function, $J_2(k \rho z)$, in the profile integral in that equation.

For the small momentum case, the Taylor's series of the Bessel functions implies the 
following analytical expansion on $k \rho$,

\beq
\int_{0}^{\infty} z dz J_2(k \rho z) \phi(z) \ \propto (k \rho)^2 
\left( 1 + \lambda_p (k \rho)^2 + \dots
\right) \ ,
\label{J2pow}
\eeq 
and we will then obtain

\beq
\alpha(k) \ = \ \frac {k^4}{18 \pi n} \ \left( 1 \ + \ \frac {\delta \rho^2}{\rhob^2} \ 
\left( 48 + 48 \lambda_p k^2 \rhob^2 + {\cal O}\left( k^4 \rhob^4 \right) \right)
 \ + \ {\cal O}\left( \frac {\delta \rho^{4}}{\rhob^{4}}\right) \right) \ ,
\label{alpha-k4dev}
\eeq 
where, up to that order, only the coefficient $\lambda_p$ keeps the profile function 
information. That coefficient is furthermore a negative one because of the alternance of signs in 
the Taylor expansion of $J_2$,

\beq
J_2(x) \simeq \frac {x^2} 8 - \frac {x^4} {16} + \dots \ ,
\eeq
and the correction to the $k^4$-power law is hence negative.
However, this result is only general provided the analyticity for the $k \rho$ expansion in  
\eq{J2pow} and, for instance, the BPST profile disregards this condition as can be seen by just  
expanding \eq{fBPST} on $k \rhob$,

\beq
\rhob f(k \rhob) \ = \ 
 \left\{ - \frac 1 8 + 2 \Gamma_E + \frac 1 2 \ln\left( \frac {k \rhob} 2 \right) \right\} \
k^2 \rhob^2 \ + \ \dots \ ,
\eeq
which is clearly non analytical in $k \rhob = 0$. Nevertheless, we show in plot \ref{Xplot} that 
$f(k \rhob)$ is negative for all $k \rhob$ and the correction to the $k^4$-power law in \eq{alpha-k4ex} 
obeys to the same pattern as the one in \eq{alpha-k4dev} with negative $\lambda_p$. The reason for this 
behaviour of $f$ can be mainly found in the large momentum limit.

For the large momentum case, the oscillating nature of $J_2$ leads the asymptotical behaviour of 
the profile integral in \eq{J2pow} to be dominated by the condition $\phi(0)=1$ and hence by 

\beq
\lim_{a \to 0} \int_{0}^{\infty} z dz J_2(k \rho z) \ e^{-a z} \ = \  \frac 2 {(k \rho)^2} \ .
\label{2k-2}
\eeq 
Thus, for $k \rho >> 1$,

\beq
\rhob f(k \rhob) \ = \ \left. \rhob \frac d {d\rho^2} \ln\left( \frac 2 {(k \rho)^2} \right) 
\right|_{\rho=\rhob}\ = \ -2 \ .
\eeq 
As can be seen in plot \ref{Xplot}, $\rhob f$ for both BPST and DP-inspired profile joins the 
horizontal asymptote $-2$ as soon as $k\rhob \sim 7$. 

Of course, this is not a rigorous proof of that corrections to the $k^4$-power law should be 
negative for any instanton profile (our fitting window is around $k \rhob \simeq 1$), but leads 
us to reasonably assume that this is the case for most of the well-behaved profiles. 

\section{Some results for DP distribution of BPST instantons}
\label{B}

This appendix is devoted to present particular results for Green functions and for the 
MOM QCD coupling constant by employing the distribution for BPST instantons given in 
\eq{distr}. We will give results for both BPST and DP-inspired profile functions.

\paragraph{The BPST profile.\\ }

The use of  BPST solution, $\phi(|x|/\rho) =\rho^2/(x^2+\rho^2)$,
implies neglecting the effect of neighbouring instanton's  classical interaction. 
The IR behavior of gluon Green functions for this profile is:

\beq\label{eq24}
G^{(m)}_{(I)}(k^2)&=&\frac{k^{2-m} n}{m \ 2^{2m-2} } \left( \frac \beta 6\right)^{m/2} \
< \rho^{3m} I(k\rho)^m > \ \nonumber \\ 
&=&  n \ \frac{4 k^2}{m} \ 
\left( \sqrt{\frac \beta 6} \frac{4 \pi^2}{k^4} \right)^{m} 
 \\
&\times& \ \left\{ \begin{array}{l} 
\displaystyle 
1 +
O\left(\frac 7 {2 \overline{\rho^2} k^2} \right)  \quad {\rm for }\quad k\rho \gg 1 \\
\displaystyle 
\frac{\gamma(\frac{7+2m}{2})}{4^m \gamma(\frac{7}{2})} \ \left( 
\frac 2 7 \overline{\rho^2} k^2 \right)^m \left(
  1 + b_m(k^2) \frac 2 7 \overline{\rho^2} k^2 + 
O\left(\left( \frac 2 7 \overline{\rho^2} k^2 
\right)^2 \right) \right)  \\ 
\rule[0cm]{8.5cm}{0cm} \quad {\rm for }\quad k\rho \ll 1 \ ,
\end{array}
\right. \nonumber
\label{averm}\eeq 
where $\beta=6/g_{\rm Latt}^2$ is the lattice parameter for the bare
coupling and   

\beq\label{eq25}
b_m(k^2) \ = \ \frac{m(7+2m)}{16} \left( 2 \gamma_E-\frac 3 2 + 
\psi\left(\frac{9+2m}{2}\right) + \ln{\frac 2 7 \overline{\rho^2} k^2} 
-\ln{4} \right) \ .
\eeq
Where $\gamma_E$ the Euler constant
( 0.577216...) and $\psi(z)=\gamma'(z)/\gamma(z)$ the Euler's
``digamma'' function. The first correction to the low momentum 
$k^4$ behavior of the quantity (\ref{mom}), is then given by:

\beq
\label{alphaIRtr}
\alpha_s(k) \ = \ \frac {121}{1134 \pi n} \ k^4 \
\left( 1 + \frac 6 {56} \overline{\rho^2} k^2 \left( 2 \gamma_E + 2 
\ln{\frac{\overline{\rho^2} k^2}{56}} + \frac{22567}{3465}\right) + \dots 
\right ) \ , 
\eeq
 valid for $\overline{\rho^2} k^2 << 7/2$.

If we fit numerically the prediction of $\alpha_s(k)$ to the power formula in \eq{alpha-k(4-e)} 
(with $\overline{\rho}=\sqrt{\overline{\rho^2}}=1.5 \Gev^{-1} = 0.3$ fm from table  \ref{tabfits})
in the window  (0.3--0.9) GeV~\footnote{This window is slightly
larger  than our lattice window (0.44--0.89) GeV. The question here is the
validity of approximating  eqs~(\ref{eq24},\ref{eq25}) by a power law which
is shown to extend further than our lattice fitting window.} we will obtain:

\beq\label{alphaIR}
\alpha_s(k) 
\  \simeq \ \frac{1.52}{18 \pi n \rhob^4} \ \left( k \rhob \right)^{3.75}\ .
\eeq

\paragraph{The DP-inspired profile.\\ }

The same averages with our optimal profile
parameterization in \eq{para}  of section \ref{GFsection} lead to 
the following low-momentum ($k<<\alpha_{\rm DP}$)  prediction for Green
functions:
\beq\label{DPcorr}
G^{(m)}_{(I)}(k^2)
&=& \ n \ \left( \sqrt{\frac \beta 6} \frac{2\pi^2}{7 \alpha_{\rm DP}^2}
  \overline{\rho^2}\right)^m
\frac{\gamma(\frac 7 2 + m)}{\gamma(\frac 7 2)} \nonumber \\ 
&\times& \quad \frac{4 k^2} m \ \left( 1 +
  \mathcal{O}\left( \alpha_{\rm DP}^2 \overline{\rho^2}, k^2\overline{\rho^2}
  \right)\right) \nonumber \ . \\
\label{averm2}
\eeq

The result  equivalent to  \eq{alphaIR} for 
this profile, using the same fitting procedure over the same window,
with a phenomenological radius $\overline{\rho}=1.5$ GeV 
and $\alpha_{\rm DP} \rho = 0.675$ from table \ref{tabfits}, is:
\beq
\label{alphaIR2}
\alpha_s(k) \  \simeq  \ \frac{1.55}{18 \pi n \rhob^4} 
\ \left( k \rhob \right)^{3.83} \ .
\eeq

\section{The ratio-ansatz}
\label{C}

In ref.~\cite{Shur-SU2} the following trial function, named ratio-ansatz, 

\beq\label{r-ansatz}
g B_\mu^a(x) \ = \ \frac {\displaystyle 2 \sum_{i=I,A} R_{(i)}^{a\alpha} \overline{\eta}^\alpha_{\mu\nu} 
\frac {y_i^\nu}{y_i^2} \ \rho_i^2 \frac{f(y_i)}{y_i^2} } 
{\displaystyle 1 + \sum_{i=I,A} \rho_i^2 \frac{f(y_i)}{y_i^2}} \ ,
\eeq
where $y_i=|x-z^i|$, was proposed to avoid singularities not physically justified at the 
center of each instanton. In \eq{r-ansatz} $\overline{\eta^\alpha_{\mu\nu}}$ should be replaced by 
$\eta^\alpha_{\mu\nu}$ when suming for anti-instantons as $i=A$.
$f(x)$ is a shape function that obeys $f(0)=1$ in order not to spoil the 
field topology at the instanton centres and that provides sufficient cut-off at large distances for 
the sum convergence. One of the physical motivations of such a large distances behaviour suggested 
in~\cite{Shur-SU2} is, in fact, the work of DP~\cite{Diakonov:1983hh}. The author of~\cite{Shur-SU2}, 
for simplicity, uses the following gaussian shape:

\beq\label{Gauss}
f(x) \ = \ \exp\left(-C \frac {x^2}{\rho^2} \right) \ ,
\eeq
and obtain the coefficient $C$ by the minimization of the action per particle for some statistical 
ensemble of instanton.
However, the minimization of the repulsion of instantons in matter leads, according 
to~\cite{Diakonov:1983hh}, to the large distance behaviour given in \eq{K2}. 
Then, why not to use:

\beq
f(x) \ = \ \frac{\alpha_{\rm DP}^2 x^2}{2} \ K_2(\alpha_{\rm DP} \rho \frac x \rho) 
\sim
\left\{ \begin{array}{ll}
1 & \mbox{\rm for } x \to 0 \\
\propto \ x^{3/2} e^{-\alpha_{\rm DP} x} & \mbox{\rm for } x \to \infty
\end{array} 
\right. 
\eeq
Thus, if $y_i >> \rho_i$ for all $i$ $g B_\mu^a$ behaves as expected following DP, 

\beq
g B_\mu^a(x) \ \sim \ 2 \sum_{i=I,A} R_{(i)}^{a\alpha} \overline{\eta}^\alpha_{\mu\nu} 
\frac {y_i^\nu}{y_i^2} \ \frac {\alpha^2_{\rm DP} \rho^2} 2 K_2(\alpha_{\rm DP} y_i) \ ,
\eeq

and as
$y_i/\rho_i \to 0$ for any $i=j$ 
\beq\label{xlow}
g B_\mu^a(x) \ & \sim & \ 2 R_{(j)}^{a\alpha} \overline{\eta}^\alpha_{\mu\nu} 
\frac {y_j^\nu}{y_j^2} \ \frac {K_2(\alpha_{\rm DP} y_j)}{\displaystyle \frac 2 
{\alpha^2_{\rm DP} \rho^2}\ + \ K_2(\alpha_{\rm DP} y_j)} \nonumber \\
& \to &  
2 R_{(j)}^{a\alpha} \overline{\eta}^\alpha_{\mu\nu} 
\frac {y_j^\nu}{y_j^2} \ \frac 1 {1+\frac {y_j^2}{\rho_j^2}} \ .
\eeq
In both large and small distances regimes we obtain\footnote{it should be noticed that near one 
particular instantons, and hence far from the others, the instanton profile (last expression in the 
first line of \eq{xlow}~) is exactly the one we conjecture in this work.} 
the same through the sum-ansatz \eq{amuins} with the instanton profile \eq{para}.
The role of the coefficient $C$ in \eq{Gauss} is played by $\alpha_{\rm DP} \rhob$, and the 
minimization of the action per particle in~\cite{Shur-SU2} leading to the determination of $C$ 
is acomplished by \eq{good} resulting from the DP variational procedure. Of course, we do not  
claim that the sum-ansatz with the profile \eq{para} leads to exactly the same results as the 
ratio-ansatz approach because the latter would imply that distances other than low and large ones have 
no influence on the Green functions. However, since the low momentum Fourier transform of the gauge 
field is dominated by large distances, at least the low-momentum behaviour of Green functions can 
be effectively described by the independent-pseudoparticle sum-ansatz approach.

\end{document}